# New peculiarity in the temperature and size dependence of electron-lattice energy exchange in metal nanoparticles

**Petro Tomchuk**[*,1], **Yevgen Bilotsky**[**,2]

[1] Institute of Physics, N.A.S. of Ukraine, UA-03028 Kyiv, Ukraine
[2] Aalto University Foundation School of Chemical Technology,
Department of Material Science and Engineering,
P.O. Box 16200, FIN-00076 AALTO, Finland

* e-mail ptomchuk@iop.kiev.ua,
** Corresponding author: e-mail yevgen.bilotsky@aalto.fi

Abstract- The work is dedicated to the development of the electron-lattice energy exchange theory in metal particles and it dependence of on particle size and electron temperature. We obtained the expressions for the constants of the electron-lattice energy exchange using the quantum-kinetic approach for infinite metal, and compared it with the result of classical kinetic approach. Both methods give the same result. But for the metal particles smaller than the mean free path of electrons, as we show, a quantum-kinetic approach (in its standard form) can not be applied,  while the classical approach can be easily generalized to the particle of such sizes.

We found a new peculiarity of the electron-lattice energy exchange -  its temperature dependence around certain size - "magic size" of metal nanoparticles. Our calculation shown, that this  changes of the electron-lattice energy exchange are associated with alteration of acoustic subzones which takes part in energy transfer. The change of the electron temperature can also lead to changes (around  the "magic size") in the acoustic subzones, which are involved in energy transfer. As a result, constant of the electron-phonon interaction in nanoparticles can be either much higher or much lower than the corresponding constant for infinite metal.

## 1 INTRODUCTION

Metal nanoparticles (MN) and their ensembles have unique physical properties [1], [2] that distinguish them from bulk metals and from macromolecules. As an example, MN have been used for changing reflection [3], high photon energy absorption and energy transfer [4-5]. MN are used in bio-sensors [6]  and in genetics to visualize cells structure [7] , in medicine as an antibacterial agent [6], as well as in the treatment of cancer [8]. Some new phenomena, like thermoacoustic phenomena in MN generated by an ultrashort laser pulse [9-10], has been predicted. The crucial point in these phenomena is the appearance of hot electrons, which is associated with intensity of electron-phonon interaction. In turn, this interaction is size dependent for small metal clusters, in particular for MN. Therefore, the study of the features of this interaction is an important task.

Nowadays the whole range of metal nanoparticles, from a few atoms clusters to mega  clusters, widely studied. Physical properties of small clusters close to the properties of macromolecules, and the properties of large clusters are close to properties of infinite solids. However, when the size of nanoclusters becomes comparable with the size of  some physical parameters, such as the mean free path of electrons and phonons, skin-layer, etc., properties of a cluster change significantly. In this paper, we shall speak mainly about of clusters, with the size of the order to (or less than) the mean free path of electrons but large than the Debye wavelength of electrons.

The optical properties of  MN has been often used owing to their ability to intensively absorb and scatter light and generate strong local field near the plasmon resonance frequencies. During the absorption of intense laser radiation, electron temperatures can reach very high level. Such electrons are called the hot electrons. Hot electrons modify the optical properties of MN which leads to some nonlinear effects.

The two-temperature approximation [12] for the hot electrons phenomena description is widely used. In this model the hot electrons and phonons characterized by their own temperatures. These temperatures are usually obtained by solving the equations of energy balance. In most publications the equations for the infinite metals have been used (as it was in the original paper [12]), even for investigation of nanoclusters. Usually, these equations modified for MN by adding the surface terms only  (see the review[18]). But the bulk energy exchange depends on the size of the MN. When the MN size becomes smaller than the electron mean free path then its motion becomes ballistic (electron moves free from wall to wall), and this is manifested in the energy exchange [13]. In this case the constant of electron-phonon interaction oscillates as a function of the MN size and disappear if this size is smaller than the critical size. For such size only the surface energy exchange remains [14].

In [12] has been shown (for the infinite metal) how the electron-phonon energy exchange can be described in kinetically and classical approaches.

In the next section the expression of the electron-phonon energy exchange is delivered by using the quantum kinetic approach. We found the limitation of applicability of conventional quantum kinetic method for small metal particles, when the size of nanoparticles is smaller then the mean free path of an electron, but as we have showed in [15], the classical approach can be generalized for ballistic electron's movements.

For readers convenience we presented our results from [15] in the Section 3. We also report about a new feature of the electron-phonon energy exchange in small metal particles – the limitation of the particle size only along the electron velocity is important, but not in transversal directions. We may say that the electron interacts with phonons, which are similar to phonons in thin metal layers.

The main result of this section is obtaining the electron-lattice energy exchange in the MN by taking into account the contribution of all electrons, by averaging over all electron velocity consistent to the Paul principle. As a consequence of this approach we found a new peculiarity of the electron-lattice energy exchange - its temperature dependence around certain size - "magic size" of metal nanoparticles. This changes are associated with alteration of acoustic subzones which takes part in energy transfer.

In the Discussion section we briefly discuss the physical background of found phenomena.

## 2 Quantum kinetic approach to the hot electrons in metal

The irradiation of MN or bulk metal leads to heating of the electron gas first. Due to electron-lattice interaction the metal lattice is also heated, but the electrons temperature in MN can the different from the lattice temperature. The equation of energy balance for determining the temperatures can be consistently obtained from the kinetic equations for distribution functions of electrons and phonons. However, there is the sizes of MN for which the kinetic method in a standard form is not applicable. Increasing the size of MN leads to transition to the bulk metals, where the movement of electrons of the conduction band has quasi classic character, while decreasing the size of MN leads to a quasi molecular metal clusters, in which electrons move by quantum orbits. In this article the quasi molecular clusters will not be discussed. We assume that the conduction electrons in the MN can be characterized by quasi continuous momentum $\vec{p} = \hbar \vec{k}$ ($\vec{k}$ is the wave vector) and the distribution function $f_{\vec{k}}$ of probability to find the quantum state with momentum $\vec{k}$. The kinetic equation for MN, which is subjected by electromagnetic radiation, can be written as

$$\frac{\partial f_{\vec{k}}}{\partial t} + \vec{v}\frac{\partial f_{\vec{k}}}{\partial \vec{r}} + \frac{e\vec{E}_{in}}{\hbar}\frac{\partial f_{\vec{k}}}{\partial \vec{k}} + \hat{I}f_{\vec{k}} = 0, \qquad (1)$$

where $t$ is the time, $\vec{r}$ is the coordinates vector of the electron, $\vec{v}$ - its velocity ($\vec{v} = \hbar\vec{k}/m$, $m$ - electron's mass), $\vec{E}_{in}$ - electric field, induced by the external field of the electromagnetic waves in MN and $\hat{I}f_{\vec{k}}$ - collision integral. The collision integral consists of the sum of two integrals

$$\hat{I}f_{\vec{k}} = \hat{I}_{ee}f_{\vec{k}} + \hat{I}_{ph}f_{\vec{k}}, \qquad (2)$$

where $\hat{I}_{ee}f_{\vec{k}}$ is the integral of electron-electron collisions and $\hat{I}_{ph}f_{\vec{k}}$ - the integral of electron-phonon collisions.

We will consider the relaxation processes, after the electron temperature (due to intense electron-electron interaction) has been set ($\tau_{ee} \sim 10^{-14} s$), therefore, in this stage, the integral of electron-electron collisions is omitted ($\hat{I}_{ee}f_{\vec{k}} \approx 0$).

The integral of electron-phonon collisions, with the phonon wave vector $\vec{q}$, can be written in the usual form:

$$\hat{I}_{ph}f_{\vec{k}} = \sum_{\vec{k}',\vec{q}} W_{\vec{k},\vec{k}',\vec{q}} \begin{Bmatrix} \begin{bmatrix} (N_{\vec{q}}+1)f_{\vec{k}}(1-f_{\vec{k}'})- \\ -N_{\vec{q}}f_{\vec{k}'}(1-f_{\vec{k}}) \end{bmatrix} \times \\ \times \delta\begin{bmatrix} \varepsilon_{\vec{k}'} - \varepsilon_{\vec{k}} + \hbar\omega_{\vec{q}} \end{bmatrix} + \\ + \begin{bmatrix} N_{\vec{q}}f_{\vec{k}}(1-f_{\vec{k}'})- \\ -(N_{\vec{q}}+1)f_{\vec{k}'}(1-f_{\vec{k}}) \end{bmatrix} \times \\ \times \delta\begin{bmatrix} \varepsilon_{\vec{k}'} - \varepsilon_{\vec{k}} - \hbar\omega_{\vec{q}} \end{bmatrix} \end{Bmatrix}. \qquad (3)$$

Here $W_{\vec{k},\vec{k}',\vec{q}}$ is the probability of electron transition from the state $\vec{k}$ to the state $\vec{k}'$ per second, $N_{\vec{q}}$ is the distribution function of phonon energy $\hbar\omega_{\vec{q}}$. We will take $N_{\vec{q}}$ as the Planck's function:

$$N_{\vec{q}} = \left\{ \exp\left(\frac{\hbar \omega_{\vec{q}}}{\theta}\right) - 1 \right\}^{-1}, \qquad (4)$$

where $\theta$ is the phonon temperature in energy units ($\theta \equiv k_B T$, $k_B$ - Boltzmann constant and $T$ is the absolute temperature).

Based on a second quantization, the Hamiltonian of the electron-phonon interactions comes as

$$\hat{H}_{int} = \sum_{\vec{k},\vec{k}',\vec{q}} C_{\vec{k},\vec{k}',\vec{q}} \left\{ b_{\vec{q}} + b^+_{-\vec{q}} \right\} a^+_{\vec{k}} a_{\vec{k}'}, \qquad (5)$$

where $a^+_{\vec{k}}$ and $a_{\vec{k}}$ are creation and annihilation operators for an electron in the state $\vec{k}$ (and similarly for the phonon's operators $b^+_{\vec{q}}$ and $b_{\vec{q}}$), then according to quantum mechanics

$$W_{\vec{k},\vec{k}',\vec{q}} = \frac{2\pi}{\hbar} \left| C_{\vec{k},\vec{k}',\vec{q}} \right|^2. \qquad (6)$$

For obtaining the Hamiltonian of the electron-phonon interaction in the form (3) we use the Bardeen-Shockley deformation potential

$$H_{int} = -\Lambda \, div \, \vec{u} \qquad (7)$$

with $\Lambda$ - the energy constant of the deformation potential and $\vec{u}$ is the displacement vector of the lattice:

$$\vec{u} = \left(\frac{\hbar}{2 M_i \cdot N}\right)^{1/2} \sum_{\vec{q}} \frac{\vec{e}_{\vec{q}}}{\sqrt{\omega_{\vec{q}}}} \left\{ b_{\vec{q}} + b^+_{-\vec{q}} \right\} e^{i\vec{q}\vec{r}} \qquad (8)$$

where $M_i$ is the ion mass, $N$ is the number of ions in MN and $\vec{e}_{\vec{q}}$ - mutually orthogonal unit vectors. Going to the second quantization of electron variables in Eq. (7) we have:

$$\hat{H}_{int} = \int d\vec{r} \, \Psi^+(\vec{r}) H_{int} \Psi(\vec{r}),$$

$$\Psi(\vec{r}) = \frac{1}{\sqrt{V}} \sum_{\vec{k}} a_{\vec{k}} e^{i\vec{k}\vec{r}}, \qquad (9)$$

$V$ -volume of MN.

From the expressions Eqs. (7), (8) and (9) the formula (6) comes as:

$$W_{\vec{k},\vec{k}',\vec{q}} = \frac{\pi \Lambda^2}{\rho \, s \, V} q \, \delta_{\vec{k}',\vec{k}-\vec{q}} \qquad (10)$$

Here we used the relation $M_i \cdot N = \rho V$ with $\rho$ - density, and the dispersion equation $\omega_{\vec{q}} = s q$ for longitudinal phonons (the only kind of phonons which interact with electrons), and $s$ is the phonon speed. Next, we assume that under laser irradiation of MN the distribution function of electrons can be written as:

$$f_{\vec{k}} = f_0(\varepsilon_{\vec{k}}) + f^{(1)}_{\vec{k}}, \qquad (11)$$

where

$$f_0(\varepsilon_{\vec{k}}) = \frac{1}{e^{\frac{\varepsilon_{\vec{k}} - \mu}{\theta_e}} + 1}$$

is the Fermi distribution function with the effective temperature $\theta_e$. The small correction $f^{(1)}_{\vec{k}}$ to the Fermi distribution, is the iterative solution of the kinetic equation (1). Multiplying the equation (1) by the $\varepsilon_{\vec{k}}$ and summing (or by integrating) over all $\vec{k}$ we receive the equation for the electron effective temperature (the energy balance equation)

$$\frac{\partial}{\partial t}\left(C_e \, T_e\right) = div\left(K_e \vec{\nabla} T_e\right) + Q - \left(\frac{\partial \tilde{\varepsilon}}{\partial t}\right)_{e,ph} \qquad (12)$$

where $C_e$ is the heat capacity and $K_e$ - the thermal conductivity of the electron gas. The energy absorbed of MN per unit volume is $Q = \vec{E}_{in} \vec{j}$ (with current density $\vec{j} = \frac{1}{\hbar} \sum_{\vec{k}} \frac{\partial \varepsilon_{\vec{k}}}{\partial \vec{k}} f^{(1)}_{\vec{k}}$ caused by the electrical field $\vec{E}_{in}$). The expression of the electron-phonon energy exchange (12) converts (by using Eqs. (3) and (4)) to:

$$\left(\frac{\partial \tilde{\varepsilon}}{\partial t}\right)_{e,ph} = \sum_{\vec{k}} \varepsilon_{\vec{k}} \hat{I} \, f_0(\varepsilon_{\vec{k}}) =$$

$$= \sum_{\vec{k},\vec{k}',\vec{q}} W_{\vec{k},\vec{k}',\vec{q}} N_{\vec{q}} \hbar \omega_{\vec{q}} \left\{ \begin{array}{l} e^{\frac{\hbar \omega_{\vec{q}}}{\theta}} f_0(\varepsilon_{\vec{k}})(1 - f_0(\varepsilon_{\vec{k}'})) - \\ -f_0(\varepsilon_{\vec{k}'})(1 - f_0(\varepsilon_{\vec{k}})) \end{array} \right\} \times$$

$$\times \delta\left(\varepsilon_{\vec{k}'} - \varepsilon_{\vec{k}} + \hbar \omega_{\vec{q}}\right) \qquad (13)$$

Assume that the lattice and electrons temperature satisfy the inequality $\hbar \omega_{\vec{q}} \ll \theta_e, \theta$. Expanding all functions (except $\delta$ - function) as power series of $\hbar \omega_{\vec{q}}$ and using the relation $f_0(\varepsilon_{\vec{k}'}) = f_0(\varepsilon_{\vec{k}} - \hbar \omega_{\vec{q}})$, we received from Eq. (13):

$$\left(\frac{\partial \tilde{\varepsilon}}{\partial t}\right)_{e,ph} \approx \sum_{\vec{k},\vec{k}',\vec{q}} W_{\vec{k},\vec{k}',\vec{q}} \left(\frac{\theta}{\hbar \omega_{\vec{q}}}\right) \cdot (\hbar \omega_{\vec{q}})^2 \times$$

$$\times \left\{ \frac{f_0(\varepsilon_{\vec{k}})(1 - f_0(\varepsilon_{\vec{k}}))}{\theta} + \frac{\partial f_0(\varepsilon_{\vec{k}})}{\partial \varepsilon_{\vec{k}}} \right\} \cdot \delta\left(\varepsilon_{\vec{k}'} - \varepsilon_{\vec{k}} + \hbar \omega_{\vec{q}}\right).$$

(14)

Then take into account the expression (10) and the identity

$$f_0(\varepsilon_{\vec{k}})(1 - f_0(\varepsilon_{\vec{k}})) = -\theta_e \frac{df_0(\varepsilon_{\vec{k}})}{d\varepsilon_{\vec{k}}}$$

we receive:

$$\left(\frac{\partial \tilde{\varepsilon}}{\partial t}\right)_{e,ph} = \frac{\pi \hbar \Lambda^2}{V\rho} \times$$

$$\times \sum_{\vec{k},\vec{q}} q^2 (\theta_e - \theta)\left(-\frac{\partial f_0(\varepsilon_{\vec{k}})}{\partial \varepsilon_{\vec{k}}}\right)\delta\left(\varepsilon_{\vec{k}-\vec{q}} - \varepsilon_{\vec{k}} + \hbar\omega_{\vec{q}}\right).$$

(15)

Until now we did not explore the condition that the domain is bounded, and hence phonons and electrons spectra are discrete.

For infinite metal's domains we can replace the sums by the integrals in Eq. (15)

$$\sum_{\vec{q}} \to \frac{V}{(2\pi)^3}\int d\vec{q}, \quad \sum_{\vec{k}} \to \frac{2V}{(2\pi)^3}\int d\vec{k}.$$

Ignoring $\hbar\omega_{\vec{q}}$ in the argument of $\delta$-function in Eq. (15) (elastic approximation) we come to the expression:

$$\sum_{\vec{q}} q^2 \delta\left(\varepsilon_{\vec{k}-\vec{q}} - \varepsilon_{\vec{k}}\right) \to \frac{V}{(2\pi)^3}\int d\vec{q}\, q^2 \delta\left(\varepsilon_{\vec{k}-\vec{q}} - \varepsilon_{\vec{k}}\right) =$$

$$= \frac{V}{(2\pi)^3}\frac{2\pi m}{\hbar^2 k}\frac{q_{max}^4}{4}$$

(16)

with

$$q_{max} = \begin{cases} q_D, & \text{if } q_D \leq 2k_F \\ 2k_F, & \text{if } q_D > 2k_F \end{cases},$$

(17)

where $q_D$ is the wave vector corresponding to the Debye frequency and $k_F$ is the Fermi wave vector of the electron. On applying well knowing approximation

$$-\frac{\partial}{\partial \varepsilon_{\vec{k}}}f_0(\varepsilon_{\vec{k}}) \approx \delta\left(\varepsilon_{\vec{k}} - \mu\right)$$

(18)

and using the expression (18) we transfer Eq. (15) into the form

$$\left(\frac{\partial \tilde{\varepsilon}}{\partial t}\right)_{e,ph} = (\theta_e - \theta)\frac{m^2 \Lambda^2}{2\hbar^3}\frac{V}{(2\pi)^3 \rho} \times$$

$$\times \begin{cases} q_D^4; & q_D \leq 2k_F \\ (2k_F)^4, & q_D > 2k_F \end{cases}$$

(19)

This equation coincides with Eq. (9) of [12] if $q_{max} = q_D$ and $V = 1$, with $q_D = k_\beta T_D/\hbar s$, where $T_D$ is the Debye temperature. The equation for the electron-lattice energy exchange (19) is valid for massive metal for temperatures $\theta > \hbar\omega_D$.

The classic approach for expression of the intensity of the electron-lattice energy exchange can be generalized in the case, when the motion of electrons is in the ballistic one (when the cluster size becomes smaller than the mean free path of an electron). It worth to notice that in this regime we can't use the equation the quantum kinetic (3) in present form, because for the derivation of this equation, we used the approximation

$$\left|\frac{e^{\frac{i}{\hbar}\Delta\varepsilon t} - 1}{\Delta\varepsilon/\hbar}\right|^2 = \lim_{t\to\infty}\left(\frac{\sin\left(\frac{\Delta\varepsilon t}{2\hbar}\right)}{\Delta\varepsilon/2\hbar}\right)^2 = 2\pi\hbar t\delta(\Delta\varepsilon),$$

(20)

$$\Delta\varepsilon = \varepsilon_{\vec{k}\pm\vec{q}} - \varepsilon_{\vec{k}} \pm \hbar\omega_{\vec{q}}.$$

(21)

The Eq. (20) is valid if

$$\frac{\Delta\varepsilon}{\hbar}\tau \gg 1$$

(22)

where $\tau$ is the electron relaxation time. The feasibility to use this delta function in Eq. (3) allows us to speak about the law of conservation of energy (in the energy exchange between electrons and phonons). In this case, the electron can transfer the energy only less or equal to the maximum phonon energy – Debye energy. But, when the size of nanoparticles is smaller then the mean free path of an electron, one shall use the electron passage time $\tau_{ballistic} \simeq \frac{L}{v} \approx \frac{L}{v_F}$ (from wall to wall) instead of $\tau$ in the inequality (23). While the cluster size is decreasing, the ratio

$$\max\frac{\Delta\varepsilon}{\hbar}\tau = \frac{\max \Delta\varepsilon}{\hbar}\tau_{ballistic} \simeq \frac{\hbar\omega_D}{\hbar}\frac{L}{v_F} \approx 1 \quad (23)$$

can be achieved. This means that Eq. (20) not valid anymore and we can't use delta function in Eq. (3). Therefore, there are clusters sizes for which the quantum kinetic equation can not be applied in its conventional form.

### 3 Classical description of electron-lattice energy exchange in confined space

In the previous section we obtained the expression for the electron-lattice energy exchange (19) in the kinetic equation approach. In this section we solve the same problem in the classical approach. The classic approach proposed in [16], [12] for infinite metal and has been generalized in our publications [15], [17], [13] for MN with ballistic electrons motions. Electrons in the MN are ballistic if the size of MN is smaller then the mean free electron path. Comparing the results of kinetic and classical approaches, we can trace, in which conditions the results coincide, and that brings what is new in the energy exchange for the ballistic motion of electrons (from one potential wall to the opposite one). In the classical description of electron-lattice energy exchange we start from the equation for the lattice displacement vector associated with the longitudinal acoustic lattice vibrations generated by moving electrons [17]

$$\frac{\partial^2}{\partial t^2}\vec{u} - s^2 \Delta \vec{u} = -\frac{\Lambda}{\rho}\vec{\nabla}\delta(\vec{r}_0(t) - \vec{R}) \qquad (24)$$

where $\vec{r}_0(t)$ is the electron trajectory. It is convenient to use scalar $\chi$ instead of vector $\vec{u}$ by the relation

$$\vec{u} = \vec{\nabla}\chi. \qquad (25)$$

From Eqs. (24) and (25) we obtain the equation for $\chi$:

$$\frac{\partial^2}{\partial t^2}\chi - s^2 \Delta \chi = -\frac{\Lambda}{\rho}\delta(\vec{r}_0(t) - \vec{R}). \qquad (26)$$

Since the right side of Eq. (26) is the force that generates the longitudinal acoustic lattice vibrations, then the energy transferring from an electron to the lattice per seconds is:

$$\frac{\partial \xi}{\partial t} = \Lambda \int \frac{\partial \vec{u}}{\partial t}\vec{\nabla}\delta(\vec{R} - \vec{r}_0(t))d^3R =$$
$$= \Lambda \int \left(\vec{\nabla}\frac{\partial \chi}{\partial t}\right)\vec{\nabla}\delta(\vec{R} - \vec{r}_0(t))d^3R. \qquad (27)$$

The solution of Eq. (26) gives (via Eq. (27)) the energy transmitted to the lattice. We assume that all trajectories of electrons which pass near periodic orbit between potential walls (with the distance between them $L$) will remain near it. Choose axis $Z$ along the velocity vector $\upsilon$. Then the trajectory of an electron can be written as:

$$\vec{r}_0(t) = \{0, 0, z_0(t)\}$$
$$z_0(t) = \begin{cases} \upsilon t, & t \le \tau/2 \\ L - \upsilon \cdot (t - \tau/2), & t > \tau/2 \end{cases}, \qquad (28)$$

where $\tau = 2L/\upsilon$ is the period of electron's movement. Therefore we shall try to find the solution of Eq. (27) in Fourier series form:

$$\chi(\vec{R}_\perp, R_l; t) =$$
$$= \sum_{l=-\infty}^{\infty}\int d\vec{q}_\perp \tilde{\chi}(\vec{q}_\perp, q_l; t)\exp\{i(\vec{q}_\perp \vec{R}_\perp + q_l R_l)\}, \qquad (29)$$

$$q_l = l\frac{2\pi}{L} \qquad (30)$$

with the periodic condition

$$\chi(\vec{R}_\perp, R_l; t) = \chi(\vec{R}_\perp, R_l + L; t). \qquad (31)$$

In this Section we use $\vec{q}$ as the reciprocal vector for the Fourier expansion. Upon inserting Eq. (29) into Eq. (26) we obtain:

$$\tilde{\chi}(\vec{q}_\perp, q_l; t) = -\frac{\Lambda}{(2\pi)^2 L} \times \frac{\exp(-i\vec{q}\vec{\upsilon}_\tau t)}{(\vec{q}\vec{\upsilon})^2 - (qs)^2} \qquad (32)$$

In (32) we have denoted

$$\vec{\upsilon}_\tau = \begin{cases} \vec{\upsilon}, & t < \tau/2 \\ -\vec{\upsilon}, & t > \tau/2 \end{cases} \qquad (33)$$

and $\vec{q} = \{\vec{q}_\perp, q_l\}$, $\vec{q}\vec{\upsilon} - q_l \upsilon$.

The expression (32) has a pole and therefore the integral (29) becomes undefined. Therefore, we proceed similarly to [12], assuming that the speed of sound has a small imaginary term $s = s + is_1$ which is responsible for weak sound damping.

The corresponding integral has to be taken in the sense of principal value by using the formula

$$\lim_{\varepsilon \to 0}\frac{1}{(q_l \upsilon)^2 - q^2(s + is_1)^2} \xrightarrow{s_1 \to 0}$$
$$\to P\frac{1}{(q_l \upsilon)^2 - (qs)^2} + i\pi \delta\{(q_l \upsilon)^2 - (qs)^2\} \qquad (34)$$

Note that in Eq. (34) we used the complex variables but only the real part of Eq. (27) $\mathrm{Re}\left(\frac{\partial \xi}{\partial t}\right)$ has physical meaning. By substituting the expression (32) into Eq. (27) and applying the formula (34), with the expression $\exp\{i q_l z_0(t)\} = \exp\{i q_l \upsilon_\tau t\}$, we get

$$\mathrm{Re}\left(\frac{\partial \xi}{\partial t}\right) =$$
$$= \frac{\Lambda^2 \upsilon}{\rho L}\sum_{l=1}^{\infty} q_l \int_0^{q_{max}} dq_\perp q_\perp (q_\perp^2 + q_l^2)\delta\{(q_l \upsilon)^2 - (qs)^2\}, \qquad (35)$$

where $q_{max}$ is the maximum of $q_\perp$, which corresponds to the Debye wave vector $q_D$ ($q_{max} = q_D = \frac{\omega_D}{s}$). After calculation of the integral in Eq. (35) we receive (see the details in [17])

$$\text{Re}\left(\frac{\partial \xi}{\partial t}\right) = \frac{\Lambda^2}{2\rho L \upsilon}\left(\frac{\upsilon}{s}\right)^4 \sum_{l=1}^{l_{max}} q_l^3. \quad (36)$$

The maximum value of $l_{max}$ is determined from the condition that the argument of the $\delta$ function is zero if $q_\perp \leq q_{max}$, i.e.

$$\left(\frac{\upsilon^2}{s^2}-1\right)q_l^2 \approx \frac{\upsilon^2}{s^2}q_l^2 \leq q_{max}^2. \quad (37)$$

This condition and Eq. (30) gives

$$l_{max} \leq L\frac{q_{max}}{2\pi}\frac{s}{\upsilon}. \quad (38)$$

The inequality $l_{max} < 1$ means that the argument of the $\delta$ function has no zeros. From this it follows that there will be no energy exchange between electrons and phonons, except the surface scattering. It is convenient to rewrite Eq. (38) in the form

$$l_{max} \leq \frac{L}{L_c(\upsilon)}, \quad (39)$$

where we used the definition $L_c(\upsilon) \equiv \frac{2\pi}{q_{max}}\frac{\upsilon}{s}$. In this stage, this threshold size $L_c(\upsilon)$ is the function of the velocity of interacting electron. Using the function floor $x$, which returns the largest integer not greater than $x$, we can rewrite Eq. (39) as

$$l_{max} = \text{floor}(L/L_c(\upsilon)). \quad (40)$$

The sum in Eq. (36) can be easily calculated

$$\sum_{l=1}^{l_{max}} q_l^3 \xrightarrow{q_l=l\frac{2\pi}{L}} \left(\frac{2\pi}{L}\right)^3 \sum_{l=1}^{l_{max}} l^3 = \left(\frac{2\pi}{L}\right)^3 \frac{l_{max}^2}{4}(1+l_{max})^2. \quad (41)$$

Therefore the expression (36) is

$$\text{Re}\left(\frac{\partial \xi}{\partial t}\right) = \frac{\Lambda^2 q_{max}^4}{16\pi \rho \upsilon}\left[\text{floor}(L/L_c(\upsilon))\right]^2 \times \\ \times \left[1+\text{floor}(L/L_c(\upsilon))\right]^2. \quad (42)$$

The result of for infinite metal (see [12]) comes from Eq. (42),

$$\text{Re}\left(\frac{\partial \xi}{\partial t}\right)_\infty \equiv \lim_{L\to\infty}\text{Re}\left(\frac{\partial \xi}{\partial t}\right) = \frac{\Lambda^2 q_{max}^4}{16\pi\rho\upsilon}. \quad (43)$$

The function

$$Q(x) = \frac{\left[\text{floor}(x)\right]^2}{x^4}\left[1+\text{floor}(x)\right]^2 \quad (44)$$

has quasi-oscillatory dependence on $x = L/L_c(\upsilon)$. Figure 1 shows the dependence of $Q(x)$ on the dimensionless argument $x = L/L_c(\upsilon)$.

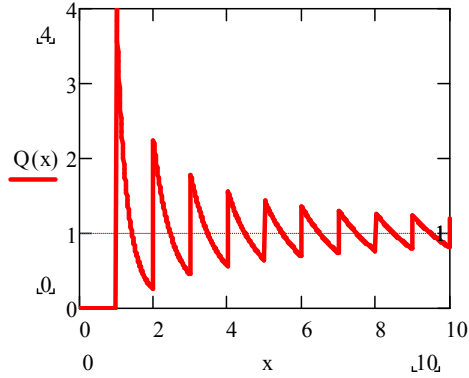

**Figure 1** The chart shows the dependence of $Q(x)$ on the dimensionless parameter $x = L/L_c(\upsilon)$ [17].

Each peak of the curve occurs due to the inclusion of the new acoustic mode in the electron-lattice energy exchange, which appears for particular sizes of MN. We can call such sizes as "magic size". The expression (43) determines the power transfers to the lattice, when the electron moving at a speed of $\upsilon$. The contribution of all electrons (per unit volume of the MN) to the electron-lattice energy exchange, which is allowed by the Pauli principle, comes by averaging of Eq. (42) with Fermi function

$$\left(\frac{\partial \tilde{\varepsilon}}{\partial t}\right)_{e,ph} = \int_{\mu-\theta_e}^{\mu+\theta_e} d\varepsilon \left\{2g(\varepsilon)f_0(\varepsilon)\text{Re}\frac{\partial \xi}{\partial t}\right\} \quad (45)$$

where

$$g(\varepsilon) = \frac{(2m)^{3/2}}{(2\pi)^2 \hbar^3}\sqrt{\varepsilon} \quad (46)$$

is the density of states. The spin of an electron gives the factor 2. In infinite metal domain, according to Eq. (43)

$$\lim_{L\to\infty}\text{Re}\left(\frac{\partial \xi}{\partial t}\right) = \frac{\Lambda^2 q_D^4}{16\pi\rho\upsilon} \quad (47)$$

and then from Eqs. (46) and (47) we get:

$$\left(\frac{\partial \tilde{\varepsilon}}{\partial t}\right)_{e,ph} \approx \theta_e \frac{m^2 \Lambda^2}{2(2\pi\hbar)^3}\frac{q_D^4}{\rho}. \quad (48)$$

This expression determines the hot electrons power losses through generating of the lattice acoustic vibrations. But apart from excitation of acoustic waves, electrons also absorb acoustic energy. In the thermodynamic equilibrium state, the energy of acoustic waves generation by electrons is equal to the energy of acoustic waves absorbed by electrons. Therefore, taking into account both effects

(generation and absorption) in Eq. (48) instead of $\theta_e$ only we shall write $\theta_e - \theta$ and that giving the result

$$\left(\frac{\partial \tilde{\varepsilon}}{\partial t}\right)_{e,ph} \approx (\theta_e - \theta)\frac{m^2 \Lambda^2}{2(2\pi\hbar)^3}\frac{q_D^4}{\rho}. \quad (49)$$

Comparing Eq. (49) with Eq. (19) (if $q_D < 2k_F$) we see that for infinite metals domain, the same results come as from classical approach and from quantum-kinetic as well (for $\theta \gg \hbar\omega_D$). In the case of finite size MN, we have:

$$\left(\frac{\partial \tilde{\varepsilon}}{\partial t}\right)_{e,ph} = (\theta_e - \theta)\frac{m^2 \Lambda^2}{2(2\pi\hbar)^3}\frac{q_D^2}{\rho} I(x,\theta_e) \quad (50)$$

where

$$I(x,\theta_e) = \int_{-1}^{1} du\, f_0(u)\, Q\!\left(x\cdot\left(1+\frac{\theta_e}{\mu}u\right)^{-1/2}\right). \quad (51)$$

In this integral we used the new variables

$$u = (\varepsilon - \mu)/\theta_e,$$
$$x = L/L_c(\upsilon_F)$$

and the substitution

$$\upsilon \to \upsilon_F\left(1+\frac{\theta_e}{\mu}u\right)^{1/2}.$$

It should be noticed that while irregularity of the function $Q(x)$ (from Eq. (44)) relates to one electron with its own velocity $\upsilon$ (see Fig. 1), the averaging over the electron velocity distribution with the function $f_0(u)$ in Eq. (51) creates a picture shown in Fig. 2. This peculiarity of electron-lattice energy exchange caused by changing the number of longitudinal phonon modes interacting with electrons (which falls into the interval $-1 \leq u \leq 1$), as the function of the electron temperature, for the given value of $L_c(\upsilon_F)$. To demonstrate the peculiarity of such dependence we start with two particular values of $L/L_c(\upsilon_F) = 0.9$ and $L/L_c(\upsilon_F) = 1.1$. According to simple approach, when the electron's velocity in Eq. (42) is fixed and equal to Fermi velocity $\upsilon_F$, $\text{Re}\left(\frac{\partial \xi}{\partial t}\right) = 0$ for $L/L_c(\upsilon_F) = 0.9$ (see Eq. (42), and Fig. 1), while Eq. (51) reveals more smoothly dependence.

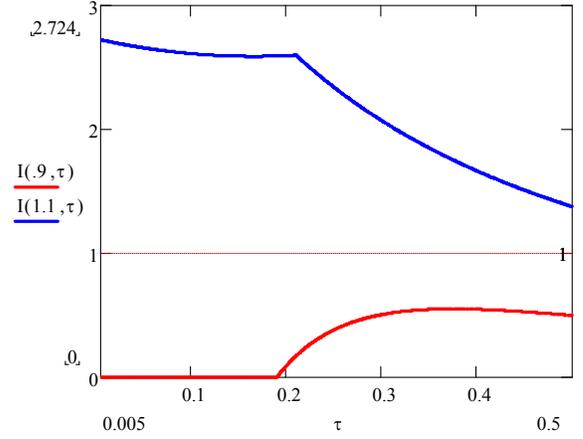

**Figure 2** The chart shows the dependence of $I(x,\theta_e)$ on the dimensionless parameter $\tau = \theta_e/\mu$ for $x = 0.9$ and $1.1$. The temperature $\theta_e$ expressed in units of $\mu$.

As it follows from Fig. 2, $I(x,\theta_e)$ and consequently $\left(\frac{\partial \tilde{\varepsilon}}{\partial t}\right)_{e,ph}$ is very sensitive to the combination of the factors of electron temperature and MN size near the "magical" size $x = 1$. As an example, for gold MN with Fermi temperature $\mu = 6.42 \cdot 10^4\, K$ and excess electron temperature on $0.08\,\mu$ this integral changes from $I(x=0.9, 0.08) = 0$ to $I(x=1.1, 0.08) = 2.667$. The difference between the integrals becomes even larger for sizes close to "magical" size, Fig 3.

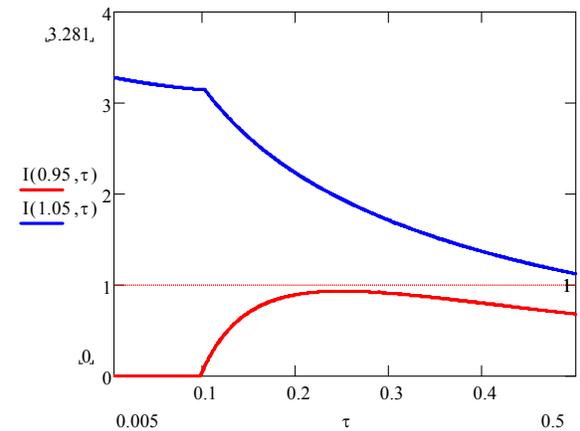

**Figure 3** The chart shows the dependence of $I(x, \theta_e)$ on the dimensionless parameter $\tau = \theta_e/\mu$ for $x = 0.95$ and $1.05$. The temperature $\theta_e$ expressed in units of $\mu$.

These features significantly complicates assessment (particularly by measuring of relaxation time of the hot electrons) of electron-phonon interactions in MN. It is the consequence of the changing the electron-phonon interactions with temperature (as above) and due to sizes distribution of MN in a set of nanoparticles which are measured. This might be one of the reason why such measurements, according to the literatures, give some times contradictive results. Some authors argue that a decrease in the size of MN increases the electron-phonon interaction, while others claim the opposite. We believe that taking into account the above features could clarify this phenomenon.

**4 Discussion**

The constant of electron-phonon energy exchange in a form in which it was received in [12] for the case of infinite metal, are often used to describe of the hot electrons in the metal nanoparticles. The goals of this article were:

1. to find out the minimum size of MN which can be used in this approach;

2. to determine corrections for the expression of the electron-phonon energy exchange, which should be made for metal nanoparticles with sizes smaller then the mean free path of electrons.

Based on the Bardeen-Schottky deformation potential (for the electron-phonon interaction), we briefly described the electron-phonon energy exchange, in quantum-kinetic and classical approaches, and shown the identity their results (in the range of applicability of both methods). The new peculiarity of the electron-phonon energy exchange in small metal particles has been found: the constraint of the particle size only along the electron velocity (but not in transversal directions) is important. Therefore, electron-phonon interaction in small metal particles has the similarity to interaction in thin metal layers.

We also showed that quantum-kinetic approach, in conventional form, becomes inapplicable for describing MN with sizes less then the mean free path of electrons, while the classical approach can be tailored for such nanoparticles. This gives a new phenomena, such as quasi-oscillations of size dependent of the bulk electron-phonon energy exchange and disappeared of this energy exchange for nanoparticles sizes less then critical one. This result can be easy understand, if consider first the solution of the Eq. (24) of acoustic lattice oscillations, generated by moving electron (with velocity $\upsilon$) in infinite lattice. The Fourier component of displacement vector comes from Eq. (24) as

$$\vec{u} = \frac{i\Lambda}{\rho \cdot (2\pi)^3} \frac{e^{-i(\omega - \upsilon t)}}{\omega^2 - q^2 s^2}, \quad (52)$$

$$\omega = \vec{q} \cdot \vec{\upsilon}.$$

This is result was obtained in [12]. As it follows from above equation, such electron generates mainly the acoustic waves, with vector $\vec{q}$ which satisfies the equation

$$\omega^2 = (\vec{q} \cdot \vec{\upsilon})^2. \quad (53)$$

From this equation and from $\vec{q} \cdot \vec{\upsilon} = q_l \upsilon$ we get

$$q_l^2 = q^2 \left(\frac{s}{\upsilon}\right)^2. \quad (54)$$

The speed of sound in metals is $s \sim 10^3 \ m/s$ and Fermi velocity is $\upsilon_F \sim 10^6 \ m/s$, as an example, in Au $s = 3.24 \times 10^3 \ m/s$ and $\upsilon_F = 1.4 \times 10^6 \ m/s$, therefore

$$\frac{q_l}{q} = \frac{s}{\upsilon} \ll 1. \quad (55)$$

From this equation we can draw two conclusions. Firstly, in an infinite metal an electron produces acoustic waves mainly in the direction perpendicular to its motion. Secondly, although the longitudinal (to the electron velocity) component of the wave vector $q_l$ is small, it still determines the resonance condition (54). And now let's see what will change in resonance condition if MN is limited (equal $L$) in the direction of electron motion. In this case

$$(q_l)_{\min} = \frac{2\pi}{L}, \quad (56)$$

therefore exist such small $L$, that

$$(q_l)_{\min}^2 = \left(\frac{2\pi}{L}\right)^2 > q_{\max}^2 \left(\frac{s}{\upsilon}\right)^2 \equiv q_D^2 \left(\frac{s}{\upsilon}\right)^2, \quad (57)$$

which means that impossible to satisfy the resonance condition (54). It should be emphasized that the condition (57) can be rewritten as

$$\frac{2\pi\upsilon}{L} > q_D s \equiv \omega_D. \quad (58)$$

This condition means, that the electron (which generates acoustic waves) oscillates with the frequency $2\pi\upsilon/L$ and this frequency is higher then maximum possible oscillation frequency (Debye frequency $\omega_D$) in the lattice. And therefore the generation of acoustic waves is impossible, and consequently the electron can not lose energy in this process. Next, we have shown that discreteness of the wave vector leads to an quasi-oscillating dependence of energy transferring from the electron to the lattice as the function of the size of MN, during the electron motion from one the surface of the nanoparticle to another. Such quasi-oscillatory electron motion determined by the dimensionless parameter

$$x = \frac{\omega_D}{2\pi\upsilon}L \equiv \frac{L}{L_c(\upsilon)}$$ (see Fig. 1). The parameter $x$ is

dependent on the electron velocity $\upsilon$ and therefore, the energy which transmitted from the electron to the lattice, shall be obtained by averaging over electron's distribution function. Most of the electrons have the energy $\varepsilon_k$ to be within

$$\mu - \theta_e \leq \varepsilon_k \leq \mu + \theta_e \qquad (59)$$

where $\mu$ is the Fermi energy and $\theta_e$ is the electron temperature, and the corresponding value of the velocity of the electrons are in the range

$$\left(1 - \frac{\theta_e}{\mu}\right)^{1/2} \upsilon_F \leq \upsilon \leq \left(1 + \frac{\theta_e}{\mu}\right)^{1/2} \upsilon_F. \qquad (60)$$

There is no peculiarity in the constant of electron-phonon energy exchange, if the peaks the function $Q$ (when $\upsilon$ changes within (60)) does not belong to this interval. Otherwise, the electron-phonon energy exchange is not a constant but is the function of the location the number of peaks inside the mentioned interval. Physical reason for this phenomenon is the change in the number of acoustic modes involved in the energy exchange (inside the interval (60)) as the function of two physical parameters – electron temperature and MN size.

**Acknowledgements** We gratefully acknowledge stimulating discussion with M. Gasik. One of the authors (PT) takes this opportunity of thanking the Fund BC 156 and another author (YB) the Helsinki University of Technology Foundation for the support for this research.